\documentclass[twocolumn,showpacs,preprintnumbers,amsmath,amssymb]{revtex4}

\usepackage{graphicx}
\usepackage{bm}

\begin{document}

\preprint{APS/123-QED}

\title{Centrifugally driven electrostatic instability in extragalactic jets}

\author{Osmanov Z.}
\affiliation{%
Georgian National Astrophysical Observatory\\ Kazbegi ave. $2^a$,
Tbilisi, 0160, Georgia
}%


\date{\today}

\begin{abstract}
The stability problem of the rotation induced electrostatic wave
in extragalactic jets is presented. Solving a set of equations
describing dynamics of a relativistic plasma flow of AGN jets, an
expression of the instability rate has been derived and analyzed
for typical values of AGNs. The growth rate was studied versus the
wave length and the inclination angle and it has been found that
the instability process is much efficient with respect to the
accretion disk evolution, indicating high efficiency of the
instability.
\end{abstract}

\maketitle

\section{Introduction}
Considering the problem of the escaping radiation from active
galactic nuclei (AGN), one has to note that we have a strong
observational evidence of a complex picture of emission spectra,
which starts from the radio, the optical up to $X$-ray and
$\gamma$-ray, with the bolometric luminosity power of AGNs lying
in the range: $\sim [10^{40}-10^{47}]erg/s$ (see Ref. 1). Origin
of emission is supposed to be energy of an accreted matter and a
spinning black hole, however, the details of the conversion
process of this energy into electromagnetic radiation is still
unknown. The discovery of blazars (see Ref. 2) as sources of
ultra-high energy radiation, has revealed that the formation and
acceleration of relativistic jets are key processes in
understanding the energy conversion mechanism.

An innermost region of AGNs is characterized by rotational motion,
and it is obvious that the rotation affects acceleration of
plasmas in jets, which consequently may influence a process of
radiation. The problem of a role of centrifugal acceleration (CA)
as the rotational effect, on a theoretical level has been studied
in Ref. 3 for the Schwarzschild black hole and it has been found
that under certain conditions the centrifugal force may change its
direction from centrifugal to centripetal. The same effect was
shown in Ref. 4 where authors studied dynamics of particles,
sliding along rotating straight channels. It has been argued that
due to the centrifugal force, the process of acceleration might be
very efficient for rotation energy extraction. On the other hand
it is obvious that no physical system can preserve rigid rotation
nearby the light cylinder surface (LCS) (an area where the linear
velocity of rotation is equal to the speed of light) where
magnetic field lines must deviate from the straight configuration.
Generalization of Ref. 4 for curved field lines has been presented
in Ref. 5 and it was found that for the Archimedes spiral a
centrifugal outflow may reach infinity, avoiding the light
cylinder problem. According to the spin paradigm (stating that, to
first order, it is the normalized black hole angular momentum,
that determines whether or not a strong radio jet is produced)
(see Ref. 6) rotation is supposed to be fundamental in formation
of outflows. Blandford \& Payne (see Ref. 7) considering the
problem of creation of the outflows, pointed out that in this
process the centrifugal acceleration may play an important role
depending on the inclination angle of magnetic field lines. A
series of works dedicated to effects of rotation, examine its role
in the nonthermal emission. One of the important examples from
this point of view is the work of Blandford \& Znajek (see Ref.
8), where they show that the rotational energy of the black hole
can be extracted electromagnetically as a Poynting flux. In the
context of AGN, Gangadhara \& Lesch (see Ref. 9) considered the
problem for understanding how efficient the rotation is for
producing $X$ - ray and $\gamma$ - ray photons. It was shown that
nearby the light cylinder surface, centrifugally accelerated
particles, due to scattering against low energetic photons, may
provide the mentioned radiation. Reconsidering the same problem
for explaining the TeV energy emission from a certain class of
blazars, in Ref. 10 it has been argued that the CA might be so
strong, that could provide ultra-relativistic electrons.

All mentioned cases exhibit the possibility of energy pumping from
rotation and in cases of the centrifugal acceleration they show
its high efficiency in the energetics of relativistic electrons in
rotating plasmas accelerating up to energies corresponding to the
Lorentz factors of the order of $10^{5-8}$ (see Ref. 10). Thus,
the amount of energy contained within a centrifugally accelerated
plasma flow is very big and if there are mechanisms for the
conversion of at least a small fraction of this energy into the
the variety of instabilities - one might find a number of
interesting consequences related to the problem of plasma energy
conversion into radiation.

Generally speaking, for the energy pumping process, one has three
fundamental stages: (a) energy conversion from rotation into
kinetic energy of the plasma background flow, (b) kinetic energy
conversion into energy of electrostatic waves and (c) conversion
of the energy of waves into radiation by means of the process of
the inverse Compton scattering (ICS) of electrostatic waves to
photons $l + e = l' + t$, which is thought to be responsible for
creation of nonthermal radiation (see Ref. 11). By Tautz \& Lerche
(see Ref. 12) the electrostatic/electromagnetic mixed mode has
been considered for relativistic jets in the context of Gamma-Ray
Bursts, to determine low-frequency instabilities. But our aim is
to consider only electrostatic waves and investigate the second
stage of the process [see (b)] and show how efficient it is.
Therefore the electrostatic wave, playing an important role in the
process of radiation might be very effective if one makes it
unstable and as we will see the centrifugal force may be very
sufficient in this context.

The centrifugally driven parametric instability has been
introduced in Ref. 13 where it was shown that the centrifugal
force acting on particles inside the pulsar magnetospheres, causes
separation of charges, leading to generation of an electrostatic
field, which excites the corresponding instability. The increment
of the linear stage has been estimated and analyzed for the Crab
pulsar and it turned out that the linear regime was extremely
efficient and short in time, indicating a need of saturation of a
growth rate. This mechanism is called parametric, because an
"external" force - the centrifugal force, playing a role of the
parameter, acts on plasma particles, changes in time and gives
rise to the instability. By applying the same approach, in Ref. 14
(to be published in Mon. Not. R. Astron. Soc.) the problem of the
reconstruction of the pulsar magnetospheres on large scales nearby
the LCS has been considered. We have studied a new parametric
mechanism and it was shown that for curvature drift waves,
corresponding instability might be extremely efficient.

In this paper a hydrodynamic approach will be used to study the
parametric mechanism of rotation induced electrostatic wave
generation in extragalactic jets, applying the approach developed
in Ref. 13 and Ref. 14 (to be published in Mon. Not. R. Astron.
Soc.).

The paper is arranged as follows. In \S\ref{sec:disper} we derive
the dispersion relation, in \S\ref{sec:discus} the corresponding
results are present and in \S\ref{sec:summary} we summarize our
results.

\section{Theory} \label{sec:disper}
%
%
%

Throughout the paper it is supposed that magnetic field lines are
straight and inclined by the angle $\alpha$ with respect to the
rotation axis. The plasma is in the frozen-in condition,
co-rotating together with the field lines with the angular
velocity $\Omega$.

It is easy to start our consideration by the local non inertial
co-rotating frame of reference. The corresponding interval in the
rigidly rotating frame will have the form (see Ref. 10):

\begin{equation}
\label{metric} ds^2 = -\left(1-\Omega_{ef}^2 r^2\right)t^2-dr^2,
\end{equation}
where $\Omega_{ef}\equiv\Omega\sin\alpha$ is the effective angular
velocity of rotation. Here we use units $c=1$. In this paper we
apply a method developed in Ref. 13 where relativistic plasma
motion was described by the single particle kinematics. According
to this approach, by using 1+1 formalism (see Ref. 15), in the
co-rotating frame of reference the equation of motion of a
particle is given by:
\begin{equation}
\label{eul_1} \frac{d {\bf p}}{d\tau}=\gamma{\bf
g}+\frac{e}{m}(\bf E+\bf v\times\bf B).
\end{equation}
where $\gamma = (1-{\bf v}^2)^{-1/2}$ is the Lorentz-factor, ${\bf
v}=d{\bf r}/d\tau$ is the velocity of the particle determined in
the 1+1 formalism, ${\bf p}\rightarrow{\bf p}/m$ - the
dimensionless momentum, $\bf g\equiv -\frac{\bf\nabla\xi}{\xi}$,
$\gamma{\bf g}$ - the effective centrifugal force (see Ref. 13)
and $\tau\equiv\xi t$ ($\xi\equiv \sqrt{1-\Omega_{ef}^2r^2} $) -
the universal time. Taking into account an identity $d/d\tau\equiv
\partial/(\xi\partial t)+(\bf v\cdot\bf\nabla)$ and a fact that Lorentz factors
in the co-rotating frame of reference and in the
inertial-laboratory frame (LF) relate to each other:
$\gamma=\xi\gamma'$ (prime denotes a physical quantity in the LF),
one can rewrite the equation of motion in the LF by following:

\begin{equation}
\label{eul} \frac{\partial{\bf p_i}}{\partial t}+({\bf
v_i\cdot\nabla)p_i}= -\gamma\xi{\bf\nabla}\xi+\frac{e}{m}\left(\bf
E+ \bf v_i\times\bf B\right),
\end{equation}
complemented by the continuity equation:

\begin{equation}
\label{cont} \frac{\partial n_i}{\partial t}+{\bf
\nabla}\cdot(n_i{\bf v_i})=0,
\end{equation}
and the Poisson equation:

\begin{equation}
\label{pois} {\bf\nabla\cdot E}=4\pi\sum_{i} e_in_{i}.
\end{equation}
$$i = e,p,b $$ here $e,p,b$ denote electrons, positrons and the
bulk components respectively. Eq. (\ref{eul}) in the zeroth
approximation (leading state), by applying the frozen-in condition
$\bf E_0+ \bf v_{0i}\times\bf B_0=0$, can be reduced to (see Ref.
4):

\begin{equation}
\label{eul_0}
\frac{d^2r}{dt^2}=\frac{\Omega_{ef}^2r}{1-\Omega_{ef}^2r^2}\left[1-\Omega_{ef}^2r^2-2
\left(\frac{dr}{dt}\right)^2\right].
\end{equation}

In Ref. 4 it has been shown that for ultra-relativistic particles
($\gamma\gg 1$) Eq. (\ref{eul_0}) has following solutions:

\begin{equation}
\label{r(t)} r(t)=\frac{V_0}{\Omega_{ef}}\sin\Omega_{ef} t,
\end{equation}
\begin{equation}
\label{vel_0} v_0(t)=V_0\cos\Omega_{ef} t,
\end{equation}
for initial conditions: $r_0 = 0$ and $V_0\sim 1$. As we see from
Eq. (\ref{eul}), the effective centrifugal force
$-\gamma\xi{\bf\nabla}\xi$ changes in time due to Eqs.
(\ref{r(t)},\ref{vel_0}) and therefore the parametric mechanism
switches on, giving rise to the electrostatic instability (as we
will see).

When studying the problem for the leading terms, plasma
oscillations have not been considered and only the effects of the
centrifugal acceleration have been taken into account (see Ref.
4). Generally speaking, different species of plasmas [electrons,
positrons and protons (bulk)] experience the centrifugal force, as
a result they separate, which will lead to the generation of the
electrostatic field, creating the Langmuir waves. This process can
be considered as a next step of the approximation. The aim is to
investigate the linear perturbation theory of the electrostatic
instability and estimate its role in relativistic AGN jets.

We start the analysis by introducing small perturbations around
the leading state:

\begin{equation}
\label{expansion} \Psi\approx \Psi^0 + \Psi^1,
\end{equation}
where $\Psi = (n,{\bf v},{\bf p},{\bf E},{\bf B})$.

Perturbing all physical quantities by following:

\begin{equation}
\label{pert} \Psi^1(t,{\bf r})\propto\Psi^1(t)
\exp\left[i\left({\bf kr} \right)\right] \,,
\end{equation}
Eqs. (\ref{eul},\ref{cont}) will get the form:

\begin{equation}
\label{eulp} \frac{\partial p^1_{i}}{\partial t}+ikv_{0}p^1_{i}=
v_{0}\Omega_{ef}^2rp^1_i +  \frac{e}{m_i}E^1,
\end{equation}
\begin{equation}
\label{contp} \frac{\partial n^1_{i}}{\partial
t}+ikv_{0}n^1_{i}+ikn_{i0}v^1_i = 0,
\end{equation}
\begin{equation}
\label{poisp} ikE^1 = 4\pi\sum_{i} e_in^1_{i},
\end{equation}
Here $E^1$ is the electric field in the LF, induced by separation
of charges.

In order to reduce the above system into a single equation let us
use an ansatz:

\begin{equation}
\label{anzp} n^1_{i}\equiv
N_{i}e^{-i\frac{V_ik}{\Omega_{ef}}\sin(\Omega_{ef} t)},
\end{equation}

then from Eqs. (\ref{eulp},\ref{contp},\ref{poisp}) one gets:

\begin{equation}
\label{eul_cont} \frac{\partial^2 N_i}{\partial t^2} =
-i\frac{e_in_i^0}{m_i\gamma_i^0}e^{iR_i}kE^1,
\end{equation}

\begin{equation}
\label{poisp1} ikE^1 = 4\pi\sum_ie_iN_ie^{-iR_i},
\end{equation}
where $R_i = \frac{V_{0i}k}{\Omega_{ef}}\sin(\Omega_{ef} t)$.

Restoring the speed of light and introducing a new variable
$N\equiv N_p-N_e$, after making the Fourier transformations, one
can easily reduce Eqs. (\ref{eul_cont},\ref{poisp1}):

\begin{equation}
\label{eq1} \omega^2N_b(\omega)=-\frac{mn_b^0\gamma_{0e}}
{2Mn_{e}^0\gamma_{0b}}\sum_s(\omega-s\Omega)^2 J_s(a)
N(\omega-s\Omega_{ef}),
\end{equation}

\begin{equation}
\label{eq2} \left(\omega^2-\frac{\omega_{e}^2}{\gamma_{0e}}\right)
N(\omega) = -\frac{\omega_{e}^2}{2\gamma_{0e}}\sum_s
J_s(a)N_b(\omega+s\Omega_{ef}),
\end{equation}
where $M$ and $m$ are masses of protons and electrons/positrons
respectively, $a\equiv ck/\Omega_{ef}$ and $\omega_{e} =
\sqrt{8\pi n^0_{e}e^2/m}$ - electron/positron plasma frequency.
For deriving Eqs. (\ref{eq1},\ref{eq2}), the identity

\begin{equation}
\label{iden} e^{\pm ix\sin\Omega_{ef} t}=\sum_s J_s(x)e^{\pm
is\Omega_{ef} t},
\end{equation}
has been used. Direct substitution of Eq. (\ref{eq1}) into Eq.
(\ref{eq2}), yields:

$$\left(\omega^2-\frac{\omega_{e}^2}{\gamma_{0e}}\right)
N(\omega)=\frac{\omega_b^2}{2\gamma_{0b}}
\sum_{s,l}J_s(a)J_l(a)\left(\frac{\omega+(s-l)\Omega_{ef}}
{\omega+s\Omega_{ef}}\right)^2\times$$
\begin{equation}
\label{disp} \times N(\omega+(s-l)\Omega_{ef}).
\end{equation}
where  $\omega_{b} = \sqrt{8\pi n^0_{b}e^2/M}$ is the plasma
frequency corresponding to the bulk component.

Generally speaking in order to solve Eq. (\ref{disp}) one has to
consider similar equations, rewriting Eq. (\ref{disp}) for
$N(\omega\pm\Omega_{ef})$, $N(\omega\pm 2\Omega_{ef})$, etc., thus
for solving the problem exactly, one needs to solve a system with
the infinite number of equations, which makes the problem
impossible to handle. Therefore the only way to overcome this
difficulty and to gain extended view of the general behavior of
the instability, is to consider physics close to the resonance
condition, which provides the cutoff of the infinite row in
Eq.(\ref{disp}) and makes the problem solvable (see Ref. 16).

If one considers in Eq. (\ref{disp}) only resonance terms,
corresponding to the following frequency $\omega_{res}\approx
\omega_{e}/\gamma_{0e}^{1/2}$, then preserving leading terms and
taking into account the resonance conditions: $\omega\approx
-s_0\Omega_{ef}$, $s_0 =  l_0\equiv
\left[\frac{\omega_e}{\gamma_{0e}^{1/2}\Omega_{ef}}\right]$ (here
$[A]$ means the integer part of $A$), the dispersion relation can
be reduced into a following single term specifying the growth rate
($\Delta\equiv \omega-\omega_{res} $) of the instability:

\begin{equation}
\label{delta1} \Delta^3\approx\frac{\omega_b^2\omega_{e}}{4
\gamma_{0b}\gamma_{0e}^{1/2}}J_{s_0}^2(a).
\end{equation}

Since we are interested in imaginary parts of $\Delta$, it is easy
to see that the following solution (see Ref. 17):

\begin{equation}
\label{delta2} \Delta\approx-\frac{1}{2}\left(1 - i\sqrt
3\right)\left[\frac{\omega_b^2\omega_{e}}{4
\gamma_{0b}\gamma_{0e}^{1/2}}J_{s_0}^2(a)\right]^{\frac{1}{3}}
\end{equation}
is responsible for the instability, the increment of which is
given by:

\begin{equation}
\label{incr1} \delta\approx\frac{\sqrt
3}{2}\left[\frac{\omega_b^2\omega_{e}}{4
\gamma_{0b}\gamma_{0e}^{1/2}}J_{s_0}^2(a)\right]^{\frac{1}{3}}.
\end{equation}

\begin{figure}
 \par\noindent
 {\begin{minipage}[t]{1.\linewidth}
 \includegraphics[width=\textwidth] {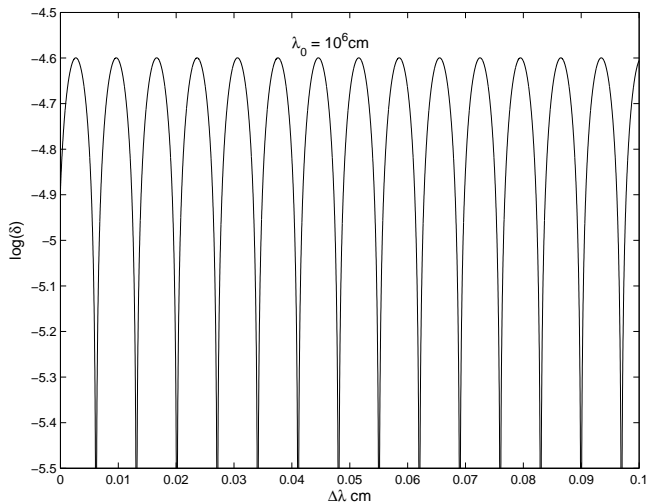}
 \end{minipage}
 }
 \caption[ ] {Dependence of logarithm of the instability rate on the wave length.
 The set of parameters is:  $\alpha =
1^{\circ}$, $M_{BH}=10^8M_{\odot}$, $\Omega = 3\times
10^{-5}s^{-1}$, $\gamma_{0b} = 20$, $\gamma_{0e} = 10^5$ and
$n^0_b = n^0_e = 0.001cm^{-3}$. As it is seen in the figure, the
increment is very sensitive on the wave length: slightly changing
$\lambda$, one may kill the instability completely.}\label{fig1}
 \end{figure}

\section{Discussion} \label{sec:discus}
%
%
%
%
%

We consider a central black hole with mass $M_{BH}=10^8M_{\odot}$
and an angular rate of rotation $\Omega = 3\times 10^{-5}s^{-1}$
of an AGN wind, making the light cylinder located at a distance
$r_L\approx 10^{15} cm$ from the center of rotation. These values
are typical for active galactic nuclei (see Ref. 18). AGN winds
and extragalactic jets are supposed to be composed of a bulk
component having the Lorentz factor of the order of $\sim 20$ and
relativistic electron-positron pairs with Lorentz factors: $\sim
10^{5}$ (see Ref. 10).

As it was already explained, the centrifugal force leads to the
generation of the Langmuir waves with the wave vector
$\overrightarrow{k}(||\overrightarrow{B})$, the growth rate of
which we are going to estimate versus the wave length $\lambda$
and the inclination angle $\alpha$ in order to investigate the
corresponding instability.

Let us suppose a jet with the opening angle $\beta = 2\alpha =
2^{\circ}$. In Fig. \ref{fig1} we show the dependence of logarithm
of the instability increment on the wave length $\lambda\equiv
\lambda_0+\Delta\lambda$, where $\lambda = 10^{6}cm$ and
$\Delta\lambda=[0,0.1]cm$.  The set of parameters is:  $\alpha =
1^{\circ}$, $M_{BH}=10^8M_{\odot}$, $\Omega = 3\times
10^{-5}s^{-1}$, $\gamma_{0b} = 20$, $\gamma_{0e} = 10^5$ and
$n^0_b = n^0_e = 0.001cm^{-3}$. The values of $n^0_b$ and $n^0_e$
are typical for extragalactic jets, which are supposed to be under
dense with respect to their ambient, obviously having density of
the order of $n_{am}\sim 1cm^{-3}$.

 \begin{figure}
 \par\noindent
 {\begin{minipage}[t]{1.\linewidth}
 \includegraphics[width=\textwidth] {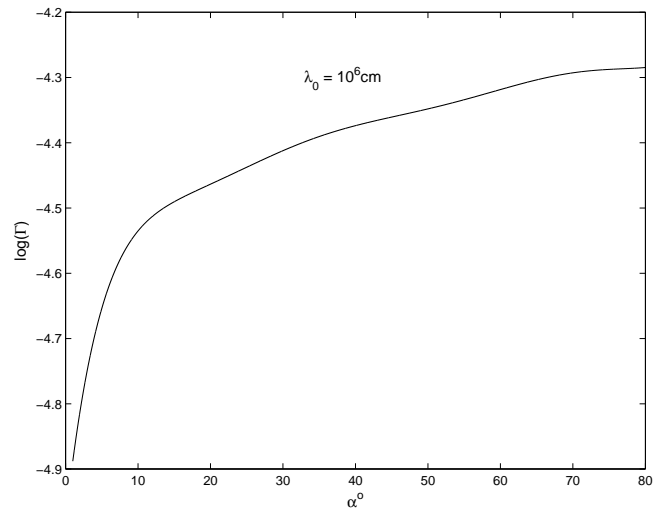}
 \end{minipage}
 }
 \caption[ ] {Dependence of logarithm of the average instability rate on the wave
 length. The set of parameters is the same as in Fig. \ref{fig1}, except
 a wider range of $\alpha$ and fixed $\lambda\equiv\lambda_0 = 10^6cm$.}\label{fig2}
 \end{figure}

An interesting feature of the result shown in Fig. \ref{fig1} is
sensitiveness of the growth rate on the wave length. As it is
clear from the plot, a small change of $\lambda$, drastically
changes the situation: for certain values of the wave length the
increment reaches its maximum level and for slightly different
values - it becomes equal to $0$. Therefore it is better to
examine an average value for each interval with a single peak,
taking into account both parameters: $\lambda$ and $\alpha$,
calculate the growth rate with respect to them:

\begin{equation}
\label{aver_delta}
{\Gamma}\equiv\frac{1}{(\lambda_2-\lambda_1)(\alpha_2-\alpha_1)}\int_{\lambda_1}^{\lambda_2}\int_{\alpha_1}^{\alpha_2}d\lambda
d\alpha\delta (\lambda, \alpha),
\end{equation}
and based on discrete data interpolate it for a wider range of
parameters. Here $\lambda_{1,2}$ and $\alpha_{1,2}$ are minimum
and maximum values of the wave length and the angle respectively
for each interval.

In spite of that jets usually have small opening angles, it is
better to study a dependence of the growth rate on inclination for
a wider range of angles, in order to understand a general
behaviour of the increment. In Fig. \ref{fig2} we show the
dependence of logarithm of the growth rate versus $\alpha$. The
set of parameters is the same as for Fig. \ref{fig1} except a
wider range of $\alpha$ and a fixed value of the wave length
$\lambda = 10^6cm$. As we see, the increment is a continuously
increasing function. This is a natural result, because when one
increases the angle, the effective angular velocity increases as
well, that makes the centrifugal force higher, leading to the more
efficient instability process.
 \begin{figure}
 \par\noindent
 {\begin{minipage}[t]{1.\linewidth}
 \includegraphics[width=\textwidth] {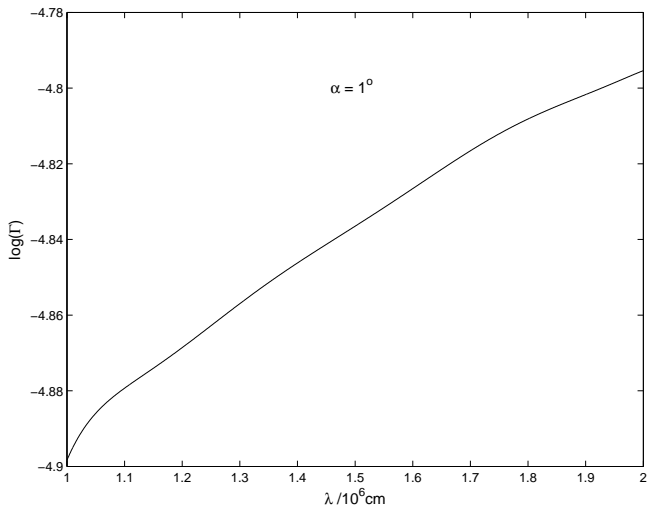}
 \end{minipage}
 }
 \caption[ ] {Dependence of logarithm of the average instability rate on the wave
 length. The set of parameters is the same as in Fig. \ref{fig1}, except
 a wider range of the wave length.}\label{fig3}
 \end{figure}

As we see from the present figure, the instability is less
efficient for smaller angles. Let us study now a dependence of the
growth rate on the wave length for the inclination angle $1^o$
when the instability is minimal (as it it is seen in Fig.
\ref{fig2}).

In Fig. \ref{fig3} the behaviour of logarithm of ${\Gamma}$ versus
$\lambda$ is shown. The set of parameters is the same as in Fig.
\ref{fig1} except a wider range of the wave length. The figure
shows that by increasing $\lambda$, the average growth rate
increases as well. Let us consider a less efficient case, thus the
smallest increment (for $\lambda = 10^6cm$) shown in the figure:
$\Gamma\sim 1.3\times 10^{-5}s^{-1}$ ($\lambda = 10^6cm $), and
see how powerful the instability is. One can easily estimate the
corresponding time scale by inversing the value of $\Gamma$:
\begin{equation}
\label{tinst}t_{inst}  =\frac{1}{\Gamma} \sim 10^5s.
\end{equation}

On the other hand jets are formed due to an accreting matter,
therefore energetics of jets is defined by the efficiency of the
accretion process, and in order to understand how efficient the
considered instability is, one has to estimate the evolution time
scale of an accreting disk and compare it with the instability
time scale.

In Ref. 19 the problem of fueling of AGNs is considered and it is
shown that for a given mass and luminosity of the AGN, mass of the
self gravitating disk can be expressed by:

$$M_{sg}  =2.76\times
10^5\left(\frac{\eta}{0.03}\right)^{-2/27}\left(\frac{\epsilon}{0.1}\right)^{-5/27}\left(\frac{L}{0.1L_E}\right)^{5/27}
$$
\begin{equation}
\label{msg}\times
\left(\frac{M_{BH}}{10^8M_{\odot}}\right)^{23/27}M_{\odot},
\end{equation}
where $\eta$ is the standard Shakura, Sunyaev viscosity parameter
(see Ref. 20), $\epsilon$ is the accretion parameter related to
the accretion mass rate $\dot{M}$ and the luminosity $L$ by the
following expression:

\begin{equation}
\label{epsilon}\epsilon\equiv \frac{L}{\dot{M}c^2},
\end{equation}
$L_E$ is the Eddington luminosity

\begin{equation}
\label{edd}L_E = 1.4\times
10^{46}\frac{M_{BH}}{10^8M_{\odot}}erg/s.
\end{equation}

Then estimating the disk's evolution time scale $t_{evol}\equiv
M_{sg}/\dot{M}$, by taking Eq. (\ref{msg}) into account, one can
easily get:

$$t_{evol}  =3.53\times
10^{13}\left(\frac{\eta}{0.03}\right)^{-2/27}\left(\frac{\epsilon}{0.1}\right)^{22/27}\left(\frac{L}{0.1L_E}\right)^{-22/27}
$$
\begin{equation}
\label{tev}\times
\left(\frac{M_{BH}}{10^8M_{\odot}}\right)^{-4/27}s.
\end{equation}

\begin{figure}
 \par\noindent
 {\begin{minipage}[t]{1.\linewidth}
 \includegraphics[width=\textwidth] {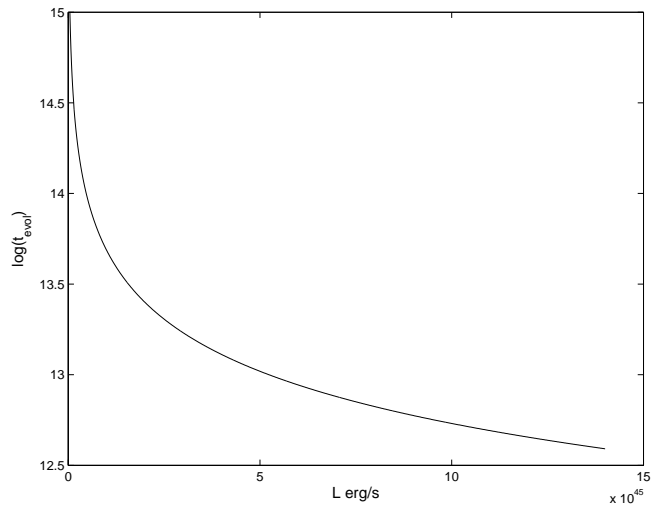}
 \end{minipage}
 }
 \caption[ ] {Dependence of logarithm of the evolution time scale on the luminosity power.
 The set of parameters is:  $\eta = 0.03$, $\epsilon = 0.1$ and $M_{BH} =
10^8M_{\odot}$. luminosity power lies in the range $L\in 1.4\times
[10^{44}; 10^{46}]erg/s$.}\label{fig4}
 \end{figure}

For taking the sense out of the considered parametric instability,
let us examine typical values of the accretion disk $\eta = 0.03$,
$\epsilon = 0.1$ and $M_{BH} = 10^8M_{\odot}$, then considering
the dependence of the evolution time scale on the luminosity, one
gets the plot shown in Fig. \ref{fig4}. For producing the plot we
have examined the following luminosity range $L/L_E\in [0.01; 1]$.
As we see in the figure, the evolution time scale is a
continuously decreasing function, and for the minimum value of the
considered luminosity, the time scale is of the order of
$10^{15}s$, whereas for the Eddington limit the time scale reaches
its minimum level $\sim 10^{13}s$.

Calculating $t_{evol}/t_{inst}$, by using Eq. (\ref{tinst}) one
can see that  the present ratio lies in the range $\sim [10^8;
10^{10}]s$, which means that the process of the instability is
much faster than typical rates of accretion, therefore the process
of conversion of rotational energy into energy of the Langmuir
waves is very rapid. This means that the linear stage has to be
short in time, and the growth rate must saturate soon due to non
linear effects, consideration of which seems to be essential.

In the introduction we have noticed that the Langmuir waves may
strongly influence processes of radiation through the ICS of these
waves against soft photons, when due to the channel $l + e = l' +
t$ the electrostatic waves lose energy, whereas the photons gain
energy. This problem was not an objective of the present paper,
which seems to be the first step for understanding the processes
of rotation energy pumping into instabilities. As a next step one
has to study this particular problem as well and see how efficient
the radiation (produced via the ICS) could be for the extremely
unstable electrostatic waves. This will be an objective of a work
which we are going to make soon or later.

\section{Summary} \label{sec:summary}
%
%
%
%

\begin{enumerate}
      \item We studied a relativistic plasma of extragalactic jets composed of
      the bulk flow and relativistic electron-positron components. The role of the
      centrifugal acceleration in generation of the parametric
      instability of the Langmuir waves has been considered.

      \item Perturbing the Euler, continuity and induction equations, preserving only
      first order terms, we have derived the dispersion relation governing the instability.

      \item Examining the resonance condition of the system, an expression of the
      growth rate has been obtained for a nearby zone of the
      light cylinder surface.

      \item Studying dependence of increment on
      the wave length for typical values of extragalactic jets, we found that the instability
      becomes efficient only for certain and narrow ranges of
      the wave length.

      \item Taking into account efficiency of an accretion process in AGNs, and
      comparing the corresponding time scale to the time
      scale of the electrostatic instability it has been found
      that even when the latter is less effective (small angles and wave
      lengths) the instability process is extremely
      efficient.

      The electrostatic instability may strongly affect processes of radiation in AGN
      jets by means of the ICS, therefore for extending this work, it is
      natural to consider a next important step and study the last stage: energy conversion
      into radiation.

      An important restriction in the present model is that we
      studied the problem for quasi-straight magnetic field lines, whereas
      in real astrophysical situations the field lines are curved
      and it is interesting to see what changes in the dynamics of
      the electrostatic wave instability, when the curvature is taken into account.

      On the other hand the curvature of magnetic field lines
      induces the curvature drift waves, which also may affect the
      process of energy pumping. In the context of pulsars
      this problem has been considered in Ref. 14 (to be published in Mon. Not. R. Astron. Soc.). It is reasonable to examine
      the same problem for relativistic AGN jets and study the role of the
      curvature drift modes in the energy conversion process.

      When studying the problem on the linear stage,
      the increment shows high efficiency of the
      electrostatic instability, therefore the non linear regime
      also has to be examined in order to study the problem for longer time scales
      when the linear approximation does not work any more. For this purpose we plan
      to implement a numerical magnetohydrodynamical code to study this particular
      case as well.

 \end{enumerate}

\section*{Acknowledgments}

I thank professor G. Machabeli for valuable discussions. The
research was supported by the Georgian National Science Foundation
grant GNSF/ST06/4-096.


\begin{thebibliography}{99}

\bibitem{catalog}  Diego F. Torres and  Sebasti\'an E. Nuza, Astrophys. J. {\bf 583},
L25 (2002)
\bibitem{hart} Hartman et al., Astrophys. J. {\bf 385}, L1
(1992)
\bibitem{abr}  M. A. Abramowicz and A. R. Prasanna, Mon. Not. R. Astron. Soc., {\bf 245},
729 (1990)
\bibitem{mrog}  G. Machabeli and A. Rogava, Phys. Rev. A {\bf 50},
98 (1994)
\bibitem{r03} A. Rogava, G. Dalakishvili and Z. Osmanov, Gen. Rel. and Grav. {\bf 35},
1133 (2003)
\bibitem{bl99} R. D. Blandford, Relativistic accretion, proc. ASP, {\bf
160} (1999)
\bibitem{bland} R. D. Blandford and D. G. Payne, Mon. Not. R. Astron. Soc., {\bf
199}, 883 (1982)
\bibitem{blzn} R. D. Blandford and R. L. Znajek, Mon. Not. R. Astron. Soc., {\bf 179},
433 (1977)


\bibitem{gan97} R. T. Gangadhara and Lesch, Astron. Astrophys., 323,
L45 (1997)
\bibitem{chemiAA} Z. Osmanov, A. Rogava and G. Bodo, Astron. Astrophys.,
470, 395O (2007)
\bibitem{schl} R. Schlickeiser, Astron. Astrophys., 410,
397 (2003)
\bibitem{incr1} G. Machabeli, Z. Osmanov and S. Mahajan, Phys. Plasmas 12,
062901 (2005)
\bibitem{totz}  R. C. Tautz and I. Lerche, Astrophys. J. {\bf 653}, 447
(2006)
\bibitem{incr2} Z. Osmanov, G. Dalakishvili and G. Machabeli,
Mon. Not. R. Astron. Soc. (2007)
\bibitem{paradigm} K. Thorne, R. Price and D. A. MacDonald, eds. Black Holes: The Membrane Paradigm (Yale University
Press, New Haven 1986) (1986)
\bibitem{silin} V. P., Silin, V. T., Tikhonchuk, J. Appl. Mech. Tech. Phys., {\bf 11}, 922 (1970)

\bibitem{mat}  M. Abramowitz and I. A. Stegun, {\it Handbook of Mathematical Functions},
Natl. Bur. Stand. Appl. Math. Ser. No. {\bf 55} (U.S. GPO, D. C.
Washington, 1965) (1965)
\bibitem{belv} G. Belvedere, L. Patern{\'o} and R. Pidatella,
Mon. Not. R. Astron. Soc., 237, 827 (1989)
\bibitem{king} A. R. King and J. E. Pringle, Mon. Not. R. Astron. Soc.,  377,
25 (2007)
\bibitem{sak} N. I. Shakura and R. A. Sunyaev, Astron. Astrophys., 24,
337 (1973)








\end{thebibliography}
\end{document}